\documentclass[journal=jpclcd,manuscript=article]{achemso}

\usepackage{chemformula} % Formula subscripts using \ch{}
\usepackage[T1]{fontenc} % Use modern font encodings

\author{Roland Kozubek}
\affiliation
{Fakult\"at f\"ur Physik and CENIDE, Universit\"at Duisburg-Essen, D-47057 Duisburg, Germany}
\author{Mukesh Tripathi}
\affiliation
{Faculty of Physics, University Vienna, A-1090 Vienna, Austria}
\author{Mahdi Ghorbani-Asl}
\affiliation
{Institute of Ion Beam Physics and Materials Research, Helmholtz-Zentrum Dresden-Rossendorf, D-01328 Dresden, Germany}
\author{Silvan Kretschmer}
\affiliation
{Institute of Ion Beam Physics and Materials Research, Helmholtz-Zentrum Dresden-Rossendorf, D-01328 Dresden, Germany}
\author{Lukas Madau{\ss}}
\affiliation
{Fakult\"at f\"ur Physik and CENIDE, Universit\"at Duisburg-Essen, D-47057 Duisburg, Germany}
\author{Erik Pollmann}
\affiliation
{Fakult\"at f\"ur Physik and CENIDE, Universit\"at Duisburg-Essen, D-47057 Duisburg, Germany}
\author{Maria O'Brien}
\affiliation
{Advanced Materials and Bioengineering Research Centre (AMBER) and School of Chemistry, Trinity College Dublin, College Green, Dublin 2, Ireland}
\author{Niall McEvoy}
\affiliation
{Advanced Materials and Bioengineering Research Centre (AMBER) and School of Chemistry, Trinity College Dublin, College Green, Dublin 2, Ireland}
\author{Ursula Ludacka}
\affiliation
{Faculty of Physics, University Vienna, A-1090 Vienna, Austria}
\author{Toma Susi}
\affiliation
{Faculty of Physics, University Vienna, A-1090 Vienna, Austria}
\author{Georg S. Duesberg}
\affiliation
{Institute of Physics, EIT 2, Faculty of Electrical Engineering and Information Technology, Universit\"at der Bundeswehr M\"unchen, D-85577 Neubiberg, Germany}
\author{Richard A. Wilhelm}
\affiliation
{Institute of Ion Beam Physics and Materials Research, Helmholtz-Zentrum Dresden-Rossendorf, D-01328 Dresden, Germany}
\alsoaffiliation
{TU Wien, Institute of Applied Physics, A-1040 Vienna, Austria}
\author{Arkady V. Krasheninnikov}
\affiliation
{Institute of Ion Beam Physics and Materials Research, Helmholtz-Zentrum Dresden-Rossendorf, D-01328 Dresden, Germany}
\alsoaffiliation
{Department of Applied Physics, Aalto University, P.O. Box 11100, FI-00076 Aalto, Finland}
\author{Jani Kotakoski}
\email{jani.kotakoski@univie.ac.at}
\affiliation
{Faculty of Physics, University Vienna, A-1090 Vienna, Austria}
\author{Marika Schleberger}
\email{marika.schleberger@uni-due.de}
\affiliation
{Fakult\"at f\"ur Physik and CENIDE, Universit\"at Duisburg-Essen, D-47057 Duisburg, Germany}

\title{Perforating freestanding molybdenum disulfide monolayers with highly charged ions}


\begin{document}

\begin{tocentry}

\includegraphics{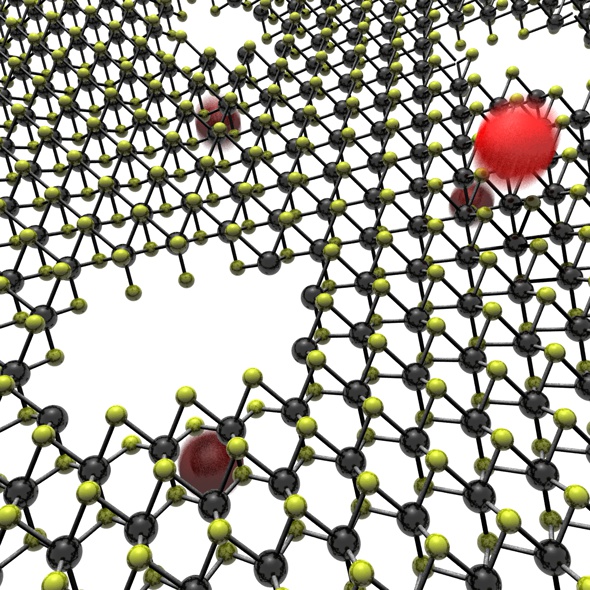}

\end{tocentry}

\begin{abstract}
    Porous single layer molybdenum disulfide (MoS$_2$) is a promising material for applications such as DNA sequencing and water desalination. In this work, we introduce irradiation with highly charged ions (HCIs) as a new technique to fabricate well-defined pores in MoS$_2$. Surprisingly, we find a linear increase of the pore creation efficiency over a broad range of potential energies. Comparison to atomistic simulations reveals the critical role of energy deposition from the ion to the material through electronic excitation in the defect creation process, and suggests an enrichment in molybdenum in the vicinity of the pore edges at least for ions with low potential energies. Analysis of the irradiated samples with atomic resolution scanning transmission electron microscopy reveals a clear dependence of the pore size on the potential energy of the projectiles, establishing irradiation with highly charged ions as an effective method to create pores with narrow size distributions and radii between ca. 0.3 and 3~nm. 
\end{abstract}

\section{Introduction}
Since the isolation of a single sheet of graphene and the discovery of its extraordinary properties, two-dimensional (2D) materials have attracted much attention \cite{Geim.2007,Lv.2015, Wang.2015}. Due to its electronic \cite{Cui.2015}, chemical \cite{Liu.2017} and optical properties \cite{Chow.2015}, molybdenum disulfide (MoS$_2$), a member of the family of 2D transition-metal dichalcogenides is one of the most prominent candidates in this field. These properties, combined with available synthesis of large-scale MoS$_2$ single layers by chemical vapour deposition (CVD) \cite{Lee.2012} open a wide range of potential applications including solar cells \cite{Singh.2017}, catalysts \cite{Yang.2014,Madau.2018}, power generators \cite{Feng.2016}, water desalination \cite{Heiranian.2015},  protein translocation \cite{Luan.2018,Chen.2018}, and DNA sequencing \cite{Sarathy.2018,Farimani.2014,Feng.2015,Smolyanitsky.2016}. 

Some of these applications require the introduction of well-defined openings with radii of a few nm in the 2D material. For example, a pore radius of $\leq$ 4 nm is suitable for the differentiation of nucleotides \cite{Feng.2015}, whereas water desalination requires pores of $\leq$ 1 nm radius \cite{Kou.2016}. For both applications the created pores would preferably have a narrow size distribution. This is typically achieved through irradiation with energetic particles, i.e.~electrons or ions \cite{Schleberger.2018}. While high-energy electrons are able to perforate freestanding MoS$_2$ easily enough, their potential in this context remains limited because mostly single or double vacancies are created \cite{Komsa.2012}. This limitation can be overcome by drilling each hole separately with a focussed beam, but this is not suitable for mass production. Another way is to expose the whole membrane to a high flux of electrons, which however will widen all defects initially present (and not only the artificially introduced ones) into larger pores and finally result in an uncontrollable size distribution and pore density as shown for single-layer hexagonal boron nitride \cite{Park.2015, Meyer.2009}. Because large-area MoS$_2$ samples are typically grown via CVD and thus contain a non-negligible number of intrinsic defects, this represents a severe drawback for up-scaling. The same is also true for methods relying on plasma treatment or chemical etching \cite{Feng.2015b}.

Individual impacts of ions with kinetic energies of a few keV have been shown to also yield mainly single and double vacancies in MoS$_2$, \cite{GhorbaniAsl.2017} and again, by focussing the beam, larger pores may be drilled into the membranes. In this way, pores with radii of 20 nm have been created \cite{Knust.2017}. For ion beams this is, however, not the limit, because in contrast to electron beams, a much wider range of experimental parameters exist for tuning the defect creation mechanism. For example, just recently it was shown that the bombardment of MoS$_2$ with swift heavy ions under grazing incidence leads to the fabrication of foldings and incisions \cite{Madau.2017}. The key feature is the energy deposition into the MoS$_2$ by electronic excitations rather than by nuclear collisions due to the very high kinetic energy of the projectile. This activates completely different defect creation mechanisms through electronic excitations resulting in structural changes far beyond single and double vacancies. The generation of a sufficiently strong electronic excitation does not necessarily require swift ions (and thus a large accelerator), but may also be achieved by using slow highly charged ions (HCIs) with velocities below the Bohr velocity $v_0$ \cite{Gruber.2016}. The deposition of their potential energy, i.e. the sum of the ionisation energies of all electrons stripped from the projectile, can result in various structural changes, ranging from hillocks \cite{Gruber.2016b} over pits \cite{Ritter.2012} in 3D materials, and holes in ultrathin carbon nanomembranes \cite{Ritter.2013}, to non-topographical features like frictional changes in graphene \cite{Hopster.2014} and hexagonal boron nitride \cite{Kozubek.2018}. Due to the energy deposition in the shallow depth range of a HCI during the interaction with surfaces, these projectiles are well suited for the modification of 2D materials. Furthermore, multiple pores can be created by exposing the whole sample to a scanned beam. This enables parallel writing and thus provides an avenue for mass production. 

In this letter, we provide experimental evidence that highly charged ions can be used to manufacture well-defined openings in freestanding MoS$_2$ with sizes suitable for applications such as DNA sequencing and water desalination. We show that the pores are produced by individual ion impacts, and determine, for a wide range of beam parameters the probability with which a pore is created. Our approach ensures a controlled pore density that is independent of any intrinsically present defects. The pore radius can be tuned with great accuracy in the range from 0.5 nm to 5 nm. In addition, we demonstrate that in particular for the smaller technologically important pores, the size distribution is extremely narrow and thus fulfills even the severe requirements of water desalination. By theoretical modelling of the ion-solid interaction, we can clearly exclude momentum transfer between the impinging ion and the atoms of the target material as the physical mechanism for pore creation. This result underlines that it is the energy associated with the charge state of the ion which gives rise to the efficient removal of atoms from a nanometer-sized area. Both the simulations as well as the experiments hint towards molybdenum enrichment at the pore edges which could be advantageous for DNA sequencing and catalysis.

%However, in order to access the full potential, a controlled modification of this new class of materials is inevitable. It is well known that surfaces of various solids as well as 2D materials experience drastic topographical changes when exposed to high energy ions\cite{Aumayr.2011, Akcoltekin.2008}. 

\section{Experiment}
For the irradiation of MoS$_2$ monolayers with HCIs, freestanding samples were used to ensure that the ions deposit their energy exclusively within the MoS$_2$, and to avoid substrate effects on the modification of the 2D material. Large-scale monolayers of MoS$_2$ were grown on SiO$_2$ via CVD in a microcavity set-up in Dublin as decribed in reference \cite{OBrien.2014} and in a custom-made process setup as an extended version of the system reported by Lee et al. \cite{Lee.2012} in Duisburg. The 2D material was transferred onto perforated membranes of amorphous carbon via wet transfer as described in reference \cite{Madau.2017}. This QUANTIFOIL\textsuperscript{\textregistered}, with a regular array of 1.2 \textmu m pores, was supported by a 400 mesh gold transmission electron microscopy (TEM) grid. This sample system allows us to exclude substrate-induced irradiation damage and to image the structure of the monolayer on the atomic scale by means of scanning transmission electron microscopy (STEM).

The irradiation of MoS$_2$ monolayers was carried out at the HICS beamline \cite{Peters.2009} at the University of Duisburg-Essen. This facility is equipped with an electron beam ion source (EBIS), which can deliver xenon ions with a maximum charge state of $q=48+$. Several charge states, ranging from $q=20+$ up to $q=40+$ were used in this work corresponding to a potential energy range of 4.6 keV $\leq E_{\textrm{pot}}\leq$ 38.5 keV. The irradiation took place under perpendicular incidence with fluences ranging from $\Phi=2.800$ \textmu m$^{-2}$ for Xe$^{40+}$ to up to $\Phi=18.000$ \textmu m$^{-2}$ for Xe$^{20+}$. To study the role of the charge state $q$, the kinetic energy was kept fixed at a constant value of $E_{\textrm{kin}}=180$ keV for all irradiations. The sample was at room temperature and the vacuum pressure during the irradiation was approximately $1\times10^{-9}$ mbar. \\
  %To minimize the impact of the ion velocity, the kinetic energy was fixed at a constant value of $E_{\textrm{kin}}=180$ keV for all irradiations.\\
\begin{figure*}
\includegraphics[width=\linewidth]{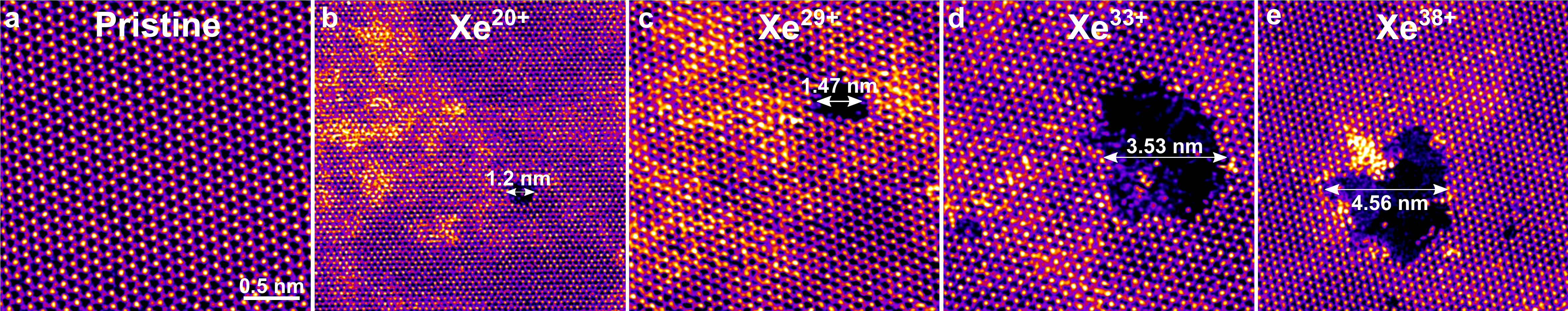}
    \caption{STEM-HAADF images, processed using the double Gaussian filter ($\sigma_1$ = 0.3, $\sigma_2$ = 0.22, weight = 0.3) \cite{Krivanek.2010} showing the MoS$_2$ lattice at atomic resolution. (a) Pristine MoS$_2$ sample. The panels show the monolayer after interaction with (b) Xe$^{20+}$, (c) Xe$^{29+}$, (d) Xe$^{33+}$ and (e) Xe$^{38+}$ ions. Pores of round shapes with a diameter in the nanometer regime are visible resulting from the energy deposition of the HCI into the 2D material. Smaller pores corresponding to one or a few missing atoms (panel e) can be created by the electron beam during imaging.}
\label{pristine}
\end{figure*} 
The STEM measurements were carried out in Vienna using an aberration-corrected Nion UltraSTEM 100 microscope with a high-angle annular dark-field (HAADF) detector. The electron acceleration voltage was 60 kV and the HAADF detector semiangle was $80-300$~mrad. A typical STEM image of a pristine MoS$_2$ monolayer is presented in Figure~\ref{pristine}a. It shows a regularly  arranged honeycomb lattice, without pores or other structural defects within the crystalline domains (the bright features are Mo, darker ones correspond to the two-atomic S column). After the interaction with HCIs, however, the STEM images reveal the presence of pores, examples of which are shown for four different charge states in Figures~\ref{pristine}b-e. Overview images are provided in the Supplementary Figure~S1.\\
%Most of the pores appear to have a round shape and show an increasing size with increasing charge state of the HCI.\\
\begin{figure*}
\includegraphics[width=\textwidth]{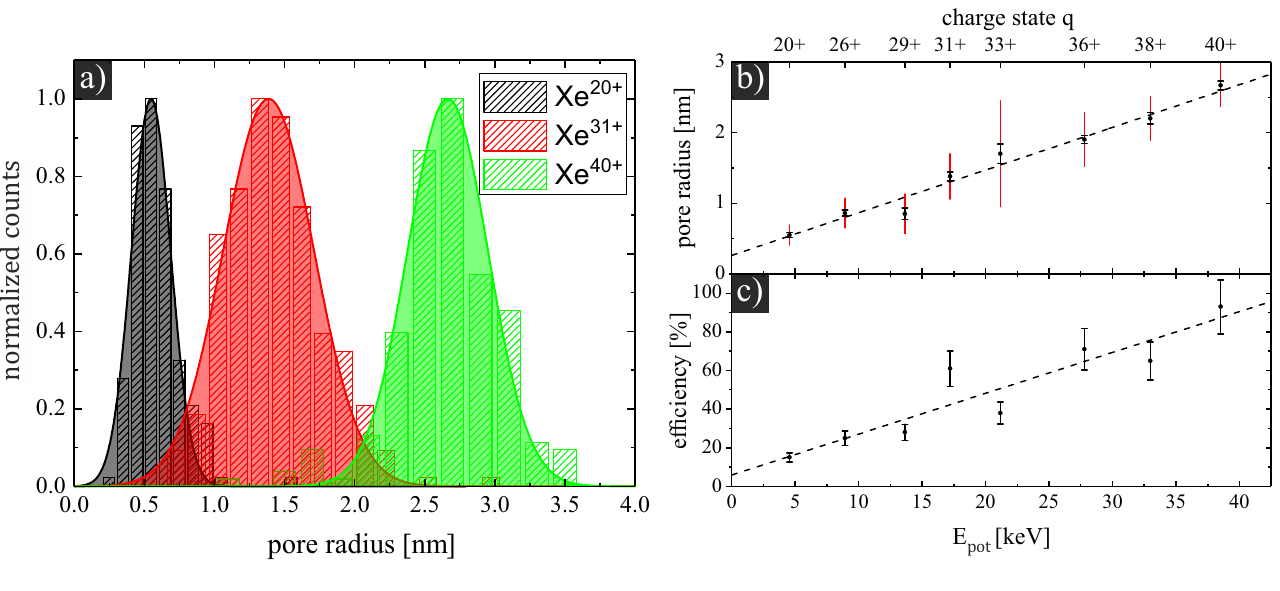}
    \caption{Distribution of pore sizes and pore creating efficiency for ions with different charge states. (a) Normalized pore size histograms for Xe$^{20+}$, Xe$^{31+}$ and Xe$^{40+}$. Each histogram is fitted using a Gaussian distribution. The diagrams in panels (b) and (c) summarize the experimental results for (b) the pore size dependency on the potential energy of the impinging Xe ion and (c) the pore creation efficiency. The red bars and error bars in panel (b) represent the standard deviation $\sigma$ of the pore size distribution and the uncertainty of the mean value ($3\cdot\sigma /\sqrt{n}$), respectively. The error bars in panel (c) correspond to the limited accuracy of the fluence determination. Both data sets were fitted using a linear function.}
\label{histogram}
\end{figure*} 

To investigate the dependency of the pore size on the charge state of the ion, an average number of over 200 pores per irradiation were surveyed and for each pore its area $A$ was determined. Since the majority of the pores shows a roughly round shape, the areas have been converted to radii $r$ using $A=\pi r^2$ and the obtained radii collected in histograms. As an example, the normalized pore size histograms for Xe$^{20+}$, Xe$^{31+}$ and Xe$^{40+}$ are shown in Figure~\ref{histogram}a. All histograms have been fitted using a Gaussian distribution.

Comparing the three histograms in Figure~\ref{histogram}a, one can clearly see that for decreasing charge state, the distribution shifts to smaller pore sizes. The mean value of the pore radius for Xe$^{20+}$ is ca. 0.55 nm, whereas pores larger than 2.65 nm in radius are created for Xe$^{40+}$. For comparing the histograms, the mean value of each distribution was determined and plotted in Figure~\ref{histogram}b. The standard deviation and the uncertainty of the mean are represented by the red bars and the error bars, respectively. This diagram summarizes the STEM data for all charge states, showing a clear correlation between the potential energy and the pore size. This dependency was best fitted using a linear function $r(E_{\textrm{pot}})=aE_{\textrm{pot}}+b$ with the slope $a=(6.04\pm0.32)\cdot10^{-2}$ nm/keV and the offset $b=(0.27\pm0.06)$ nm.

When analysing our data, it became apparent that the probability to create a pore upon impact (pore density/fluence) decreases with decreasing $E_{\textrm{pot}}$ (Figure~\ref{histogram}c). The uncertainties correspond to the limited accuracy of the current measurement in the picoampere regime. In the case of Xe$^{40+}$, an efficiency of $(93\pm14)$\% was determined, meaning that the observed pore density is nearly identical to the applied fluence. This implies that practically every HCI created a pore in the sample. However, decreasing the potential energy of the projectile, the defect creation efficiency decreases down to $(15\pm2)$\%. A linear fit of these data reveals a slope of $(2.1\pm0.4)$ \%/keV and an offset of $(6.0\pm8.3)$\%.  %This experiment however shows an constant increase of the efficiency with the potential energy.   

%A comparison to theoretical predictions using MD simulations however shows, that this value is about one order of magnitude larger than expected.

\section{Discussion}

Let us now discuss the experimental results, starting with the pore size. Figures \ref{histogram}a and b clearly show that nanometer-sized pores are induced into the 2D material due to the interaction with HCIs. The pore radius can be adjusted in the range of $0.3-3$~nm by varying the potential energy $E_{\textrm{pot}}$ of the incoming ion. Comparing the pore radius with previous experimental data of HCI-irradiated MoS$_2$ monolayers supported on KBr \cite{Hopster.2013}, it stands out immediately that the pore area is  about one order of magnitude smaller in our experiment. Although the kinetic energy of our projectiles is clearly lower (180 keV instead of 260 keV), this cannot explain the large discrepancy between the two systems. Instead, a probable reason for the reduction of the pore size is the lack of a substrate. It is well known that the substrate can have a significant influence of the modification of 2D materials \cite{Li.2015, Zhao.2015}. In addition to the direct interaction of the HCI with MoS$_2$, atoms sputtered from the substrate may lead to an additional removal of atoms from the material, resulting in an increase of the pore size for the supported sample~\cite{Kretschmer.2018}.

By decreasing $E_{\textrm{pot}}$, the radius can be reduced down to the offset value of the linear fit, as shown in Figure \ref{pristine}, representing the extrapolated limiting size of a pore that would be induced into a MoS$_2$ monolayer after interaction with a neutral Xe atom. Without any potential energy, the pore creation process would be driven by the deposition of kinetic energy. This pore of $0.27\pm0.06$ nm would correspond to ca. 9 missing molybdenum and twice as many missing sulphur atoms, and represents the lower limit of the achievable pore size in suspended MoS$_2$ by Xe irradiation for the given kinetic energy. Although theoretical simulations predicted comparable pore production after irradiation with neutral Xe projectiles \cite{GhorbaniAsl.2017}, to our knowledge, no experimental data on Xe irradiation of MoS$_2$ monolayers has been published so far.

Coming to the pore size evolution presented in Figure~\ref{histogram}b, the linear increase of the pore radius with the potential energy of the ion proves that this parameter plays a crucial role in the defect creation process. The experimental data even suggests that the number of sputtered atoms per keV of potential energy increases significantly with $E_{\textrm{pot}}$ of the HCI. While approximately 7 atoms per keV were removed with Xe$^{20+}$ (overall 33 atoms), nearly three times as many (ca. 20 atoms/keV or 765 atoms in total) have left the 2D material due to an impact with a Xe$^{40+}$. Qualitatively, similar findings have been obtained for other materials, including graphene supported by SiO$_2$ \cite{Hopster.2014}. A possible explanation could be found in the emission characteristics of the sputtered particles. For increasing potential energy, it becomes more likely to sputter clusters of atoms \cite{Schiwietz.1995}. This consumes less energy than sputtering single atoms, since the energy cost for breaking the bonds within the clusters is saved. As a consequence, more atoms (individually or in clusters) per energy unit can be emitted from the surface for higher charge states. 

Next, let us try to shed light on the defect creation process. This is driven by the energy $E^{\textrm{dep}}$, which is deposited into MoS$_2$ during the HCI-MoS$_2$ interaction. This energy can be described as
\begin{equation*}
E^{\textrm{dep}}(q)=\left(\frac{dE}{dx}\right)_n(q) + \left(\frac{dE}{dx}\right)_e(q) + E_{\textrm{pot}} ^{\textrm{dep}}(q) ,
\end{equation*}
with $E_{\textrm{pot}} ^{\textrm{dep}}(q)$ being the amount of potential energy of the HCI deposited into the electronic subsystem of the target and $\left(\frac{dE}{dx}\right)_{n,e}(q)$ being the nuclear and electronic stopping power. Whereas the dependency of $E_{\textrm{pot}} ^{\textrm{dep}}$ on the charge state $q$ is quite intuitive, it has just recently been shown that the stopping power of slow HCIs in graphene shows a strong dependency on $q$ \cite{Wilhelm.2016}.

As sputtering is typically driven by nuclear stopping, it is instructive to assess the impact of the charge-dependent nuclear stopping $\left(\frac{dE}{dx}\right)_n(q)$ of the ion on the defect creation process. To this end, we carried out molecular dynamics simulations using the large-scale atomic/molecular massive parallel simulator (LAMMPs) package \cite{Plimpton.1995}. Additional details on the MD simulation are provided in the Supporting information. 

To model the ionic charge state in the ion irradiation, a charge state-dependent potential in MD simulations based on a modified electrostatic potential for nuclear energy transfer was used \cite{Wilhelm.2016}. The electronic-stopping power $\left(\frac{dE}{dx}\right)_e$ was kept constant and charge-independent in this simulation. A value of 41 eV/\AA\ was calculated from SRIM \cite{Ziegler.2010}(irradiation of bulk MoS$_2$ using 180 keV Xe) and was adapted for the monolayer. The energy brought in by the ion impact was estimated as a product of electronic stopping power and layer thickness (6.15 \AA). It was deposited within a circular area around the impact point with a radius of 10 \AA. Simulations with different physically meaningful radii gave qualitatively similar results, with the radius being within 50\% of what is presented in Figure~3 (for details, see Supplementary Figure S2). In any case, as evident from the following discussion, charge-independent electronic stopping cannot explain the experimental observations, as the deposited potential energy $E_{\textrm{pot}} ^{\textrm{dep}}(q)$ was not taken into account. \\ 
\begin{figure}
\includegraphics[width=8cm]{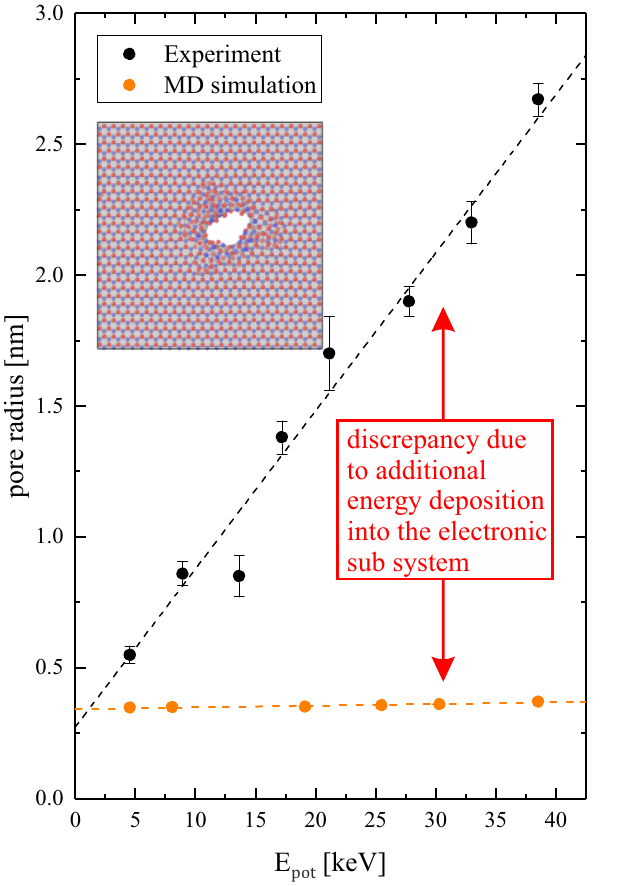}
\caption{Average pore radius in MoS$_2$ as function of the potential energy obtained by STEM measurements (black) and by MD simulation (orange). The discrepancy between the data sets decreases with decreasing amount of potential energy $E_{\textrm{pot}}$ until the extrapolated values at $E_{\textrm{pot}}=0$ keV are nearly identical. The inset shows an example of the simulated atomic structure after an impact of a Xe$^{40+}$ ion with a kinetic energy of 180 keV.}
\label{theory}
\end{figure} 
The results of the simulation (orange) together with the experimental results (black) are presented in Figure~\ref{theory}. It is first of all apparent that the MD simulation also shows a perforation of the MoS$_2$ single layer, exhibiting similar pore shapes compared to the experimental results. However, the calculated pore radii, ranging from about 0.35 nm for Xe$^{20+}$ to 0.37 nm for Xe$^{40+}$, are significantly smaller than the experimental results. Whereas for Xe$^{20+}$, the simulated pores are by a factor of ca. 1.5 smaller compared to the experiment, the discrepancy increases further to a factor of more than seven in the case of Xe$^{40+}$. 
This indicates that considering just the nuclear stopping of the projectiles to be charge-state-dependent and accounting for charge-state-independent electronic stopping is not sufficient to reproduce the experimental results. Instead, one must consider the electronic stopping to be charge-state-dependent as well, and even more importantly, implement the deposition of the potential energy of the HCI in the calculation. This, however, is not easily achieved, as e.g. not all energy stored in the projectile is deposited in the layer. For graphene, where efficient charge exchange can be expected, it has been shown that most projectiles retain a significant amount of potential energy after transmission \cite{Gruber.2016b}. Clearly, further experiments to obtain more information about the interaction of the HCIs with monolayer MoS$_2$ (i.e. energy loss, charge transfer, etc.) need to be carried out to help improving the computational models. This, however, is a major undertaking and thus beyond the scope of this paper.

The impact of an HCI on a 2D material may trigger the emission of electrons \cite{Gruber.2016b,Wilhelm.2017} as well as a local temperature increase \cite{Kozubek.2018}. In the case of a MoS$_2$ monolayer, heat treatment and electron bombardment have both been shown to result in a depletion of sulfur atoms \cite{Lin.2016,Komsa.2013}. Interestingly, indications for this are also found in our calculations. In the inset of Figure~\ref{theory}, a result of the MD simulation is depicted, showing a pore in the MoS$_2$ that is created after the impact of a Xe$^{40+}$ ion. Besides the pore creation, the simulation additionally shows amolybdenum enrichment in the direct vicinity of the pore. By comparing the amount of sputtered ions (see Supplementary Figure S3), the simulation suggests a depletion of sulfur, while sputtering of molybdenum appears to be negligible. Furthermore, the simulation suggests a displacement of molybdenum atoms from the lattice to the surface, resulting in an aggregation of molybdenum clusters at the pore edges (see Supplementary Figure S4), which has also been reported for MoS$_2$ monolayers irradiated with electrons \cite{Wang.2017, Zan.2013}. Although a direct comparison between the simulations and experiments is not warranted due to the prevailing limitations in the computational methods, we note that a similar molybdenum aggregation was in many cases also observed experimentally. For an example see Supplementary Figure~S5.

Next, let us discuss the linear evolution of the defect creation efficiency with the charge state of the ion. This linear increase without a threshold is in contrast to earlier observations with HCIs on surfaces of bulk materials and other 2D materials \cite{Aumayr.2011,Hopster.2014}. There, a threshold value for the energy deposition can be determined in a narrow range, separating a domain with a certain finite constant efficiency for the modification of the target material from the regime where the energy deposition is not sufficient to modify the irradiated system resulting in zero efficiency. In a number of experiments, a decrease of the defect creation efficiency was observed in a narrow range near the threshold value \cite{Wilhelm.2015}. In our experiment, however, the evolution of the efficiency spans an order of magnitude in potential energy. Whereas approximately every fifth ion creates a pore for $E_{pot}\approx 4.6$ keV (Xe$^{20+}$), the efficiency increases linearly up to a value close to 100\% for  $E_{pot}\approx 38.5$ keV (Xe$^{40+}$).
% JANI: This is simply not true. There is no reason why contamination would vary linearly with E_pot, and the other effects we would see in the STEM images
%This observation may be explained by a high sensitivity of the pore creation process towards various factors, which are differing for each HCI impact, such as the fraction of deposited energy as well as the the amount of contamination and the mechanical strain of the MoS$_2$ at the impact site. In addition to these considerations, the appearance of completely different types of defects like structural transformation from 2H phase to 1T phase \cite{Lin.2014,Kretschmer.2017} or the production of sulphur vacancies \cite{GhorbaniAsl.2017} could additionally explain the decrease of pore creation efficiency, since these defects are more difficult to identify in STEM and would most likely go undetected in our statistical measurements. To verify this assumption further, experimental studies need to be carried out focussing on the high technical and preparational requirements to increase the SNR in the STEM images to enable the detection of these defect structures. 
This is a truly surprising finding, which indicates that the probability for energy deposition from the HCI to the sample must depend either on the charge state or the potential energy in a manner that has not been previously observed. In a bulk sample this is obscured, because the penetration depth of the HCI is limited and thus all potential energy will be deposited close to the sample surface with 100\% likelihood. Clearly, in order to improve our understanding on the interaction between HCI and 2D materials, systematic study of HCI irradiation of structures with different properties is necessary.

Finally, we wish to point out that we see no reason why the pore creation process itself should be specific for MoS$_2$. While in other TMDCs details of the defect structure and their exact size might differ, we are confident that pore creation of similar size range is entirely feasible.

\section{Conclusion}
In summary, this paper presents the first experiments focused on the interaction of highly charged ions with freestanding monolayers of MoS$_2$. The successful creation of pores in the otherwise pristine lattice of the 2D material was observed, and the radius of the pores ranging from 0.55 nm to 2.65 nm was controlled by changing the charge state of the projectile. To disentangle the contribution of the charge dependent deposition of the kinetic energy and the deposition of the potential energy of the highly charged ion to the defect formation process, MD simulations were carried out showing that considering only the nuclear stopping to be charge-state-dependent and  electronic stopping to be charge-independent is not sufficient to reproduce the experimental results, suggesting the charge state-dependent electronic stopping and the potential energy to be the driving force.  Although additional experiments need to be carried out to clarify the underlying defect creation process, irradiation with highly charged ions clearly represents a novel technique for fabricating porous MoS$_2$ monolayers with pores in the size regime meeting the strict requirements of water desalination or DNA sequencing.

\begin{acknowledgement}

We acknowledge financial support from the DFG within the SFB 1242 ("Non-Equilibrium Dynamics of Condensed Matter in the Time Domain") Project C5, within the Project NU-TEGRAM (SCHL 384/16-1), SCHL 384/20-1, KR 4866/2-1 and WI-4691/1-1, the Austrian Science Fund projects I3181, P31605 and P28322 as well as the Wiener Wissenschafts-, Forschungs- und Technologiefonds (WWTF) project MA14-009. G.S.D acknowledges support of the Science Foundation Ireland under grant No 12/RC/2278 and 15/IA/3131. The computational support from the HZDR computing cluster is gratefully appreciated.

\end{acknowledgement}

\begin{suppinfo}

The following files are available free of charge.
\begin{itemize}
  \item Details on the molecular dynamics simulation 
  \item Figure S1: STEM overview and close-up images of pristine and ion irradiated MoS$_2$
  \item Figure S2: Results of the MD simulation, showing the pore size dependency on the energy deposition radius $r_d$
   \item Figure S3: Results of the MD simulation, showing the average amount of sputtered sulphur and molybdenum atoms.
   \item Figure S4: Results of the MD simulation, showing the atomic lattice and the molybdenum density after ion impact.
   \item Figure S5: STEM images of HCI irradiated MoS$_2$ showing a aggregation of Mo atoms on the pore edges.
\end{itemize}

\end{suppinfo}

\providecommand{\latin}[1]{#1}
\makeatletter
\providecommand{\doi}
  {\begingroup\let\do\@makeother\dospecials
  \catcode`\{=1 \catcode`\}=2 \doi@aux}
\providecommand{\doi@aux}[1]{\endgroup\texttt{#1}}
\makeatother
\providecommand*\mcitethebibliography{\thebibliography}
\csname @ifundefined\endcsname{endmcitethebibliography}
  {\let\endmcitethebibliography\endthebibliography}{}


\begin{mcitethebibliography}{52}
\providecommand*\natexlab[1]{#1}
\providecommand*\mciteSetBstSublistMode[1]{}
\providecommand*\mciteSetBstMaxWidthForm[2]{}
\providecommand*\mciteBstWouldAddEndPuncttrue
  {\def\EndOfBibitem{\unskip.}}
\providecommand*\mciteBstWouldAddEndPunctfalse
  {\let\EndOfBibitem\relax}
\providecommand*\mciteSetBstMidEndSepPunct[3]{}
\providecommand*\mciteSetBstSublistLabelBeginEnd[3]{}
\providecommand*\EndOfBibitem{}
\mciteSetBstSublistMode{f}
\mciteSetBstMaxWidthForm{subitem}{(\alph{mcitesubitemcount})}
\mciteSetBstSublistLabelBeginEnd
  {\mcitemaxwidthsubitemform\space}
  {\relax}
  {\relax}

\bibitem[Geim and Novoselov(2007)Geim, and Novoselov]{Geim.2007}
Geim,~A.~K.; Novoselov,~K.~S. The rise of graphene. \emph{Nature materials}
  \textbf{2007}, \emph{6}, 183--191\relax
\mciteBstWouldAddEndPuncttrue
\mciteSetBstMidEndSepPunct{\mcitedefaultmidpunct}
{\mcitedefaultendpunct}{\mcitedefaultseppunct}\relax
\EndOfBibitem
\bibitem[Lv \latin{et~al.}(2015)Lv, Robinson, Schaak, {Du Sun}, Sun, Mallouk,
  and Terrones]{Lv.2015}
Lv,~R.; Robinson,~J.~A.; Schaak,~R.~E.; {Du Sun},; Sun,~Y.; Mallouk,~T.~E.;
  Terrones,~M. Transition Metal Dichalcogenides and Beyond: Synthesis,
  Properties, and Applications of Single- and Few-Layer Nanosheets.
  \emph{Accounts of Chemical Research} \textbf{2015}, \emph{48}, 56--64\relax
\mciteBstWouldAddEndPuncttrue
\mciteSetBstMidEndSepPunct{\mcitedefaultmidpunct}
{\mcitedefaultendpunct}{\mcitedefaultseppunct}\relax
\EndOfBibitem
\bibitem[Wang \latin{et~al.}(2015)Wang, Wang, Wang, Wang, Yin, Xu, Huang, and
  He]{Wang.2015}
Wang,~F.; Wang,~Z.; Wang,~Q.; Wang,~F.; Yin,~L.; Xu,~K.; Huang,~Y.; He,~J.
  Synthesis, properties and applications of 2D non-graphene materials.
  \emph{Nanotechnology} \textbf{2015}, \emph{26}, 292001\relax
\mciteBstWouldAddEndPuncttrue
\mciteSetBstMidEndSepPunct{\mcitedefaultmidpunct}
{\mcitedefaultendpunct}{\mcitedefaultseppunct}\relax
\EndOfBibitem
\bibitem[Cui \latin{et~al.}(2015)Cui, Lee, Kim, Arefe, Huang, Lee, Chenet,
  Zhang, Wang, Ye, Pizzocchero, Jessen, Watanabe, Taniguchi, Muller, Low, Kim,
  and Hone]{Cui.2015}
Cui,~X.; Lee,~G.-H.; Kim,~Y.~D.; Arefe,~G.; Huang,~P.~Y.; Lee,~C.-H.;
  Chenet,~D.~A.; Zhang,~X.; Wang,~L.; Ye,~F. \latin{et~al.}  Multi-terminal
  transport measurements of MoS2 using a van der Waals heterostructure device
  platform. \emph{Nature nanotechnology} \textbf{2015}, \emph{10},
  534--540\relax
\mciteBstWouldAddEndPuncttrue
\mciteSetBstMidEndSepPunct{\mcitedefaultmidpunct}
{\mcitedefaultendpunct}{\mcitedefaultseppunct}\relax
\EndOfBibitem
\bibitem[Liu \latin{et~al.}(2017)Liu, Robertson, Li, Kuo, Darby, Muhieddine,
  Lin, Suenaga, Stamatakis, Warner, and Tsang]{Liu.2017}
Liu,~G.; Robertson,~A.~W.; Li,~M. M.-J.; Kuo,~W. C.~H.; Darby,~M.~T.;
  Muhieddine,~M.~H.; Lin,~Y.-C.; Suenaga,~K.; Stamatakis,~M.; Warner,~J.~H.
  \latin{et~al.}  MoS2 monolayer catalyst doped with isolated Co atoms for the
  hydrodeoxygenation reaction. \emph{Nature chemistry} \textbf{2017}, \emph{9},
  810--816\relax
\mciteBstWouldAddEndPuncttrue
\mciteSetBstMidEndSepPunct{\mcitedefaultmidpunct}
{\mcitedefaultendpunct}{\mcitedefaultseppunct}\relax
\EndOfBibitem
\bibitem[Chow \latin{et~al.}(2015)Chow, Jacobs-Gedrim, Gao, Lu, Yu, Terrones,
  and Koratkar]{Chow.2015}
Chow,~P.~K.; Jacobs-Gedrim,~R.~B.; Gao,~J.; Lu,~T.-M.; Yu,~B.; Terrones,~H.;
  Koratkar,~N. Defect-induced photoluminescence in monolayer semiconducting
  transition metal dichalcogenides. \emph{ACS nano} \textbf{2015}, \emph{9},
  1520--1527\relax
\mciteBstWouldAddEndPuncttrue
\mciteSetBstMidEndSepPunct{\mcitedefaultmidpunct}
{\mcitedefaultendpunct}{\mcitedefaultseppunct}\relax
\EndOfBibitem
\bibitem[Lee \latin{et~al.}(2012)Lee, Zhang, Zhang, Chang, Lin, Chang, Yu,
  Wang, Chang, Li, and Lin]{Lee.2012}
Lee,~Y.-H.; Zhang,~X.-Q.; Zhang,~W.; Chang,~M.-T.; Lin,~C.-T.; Chang,~K.-D.;
  Yu,~Y.-C.; Wang,~J. T.-W.; Chang,~C.-S.; Li,~L.-J. \latin{et~al.}  Synthesis
  of large-area MoS2 atomic layers with chemical vapor deposition.
  \emph{Advanced materials (Deerfield Beach, Fla.)} \textbf{2012}, \emph{24},
  2320--2325\relax
\mciteBstWouldAddEndPuncttrue
\mciteSetBstMidEndSepPunct{\mcitedefaultmidpunct}
{\mcitedefaultendpunct}{\mcitedefaultseppunct}\relax
\EndOfBibitem
\bibitem[Singh \latin{et~al.}(2017)Singh, Kim, Yeom, and Nalwa]{Singh.2017}
Singh,~E.; Kim,~K.~S.; Yeom,~G.~Y.; Nalwa,~H.~S. Atomically Thin-Layered
  Molybdenum Disulfide (MoS2) for Bulk-Heterojunction Solar Cells. \emph{ACS
  applied materials {\&} interfaces} \textbf{2017}, \emph{9}, 3223--3245\relax
\mciteBstWouldAddEndPuncttrue
\mciteSetBstMidEndSepPunct{\mcitedefaultmidpunct}
{\mcitedefaultendpunct}{\mcitedefaultseppunct}\relax
\EndOfBibitem
\bibitem[Yang \latin{et~al.}(2014)Yang, Fei, Ruan, Xiang, and Tour]{Yang.2014}
Yang,~Y.; Fei,~H.; Ruan,~G.; Xiang,~C.; Tour,~J.~M. Edge-oriented MoS2
  nanoporous films as flexible electrodes for hydrogen evolution reactions and
  supercapacitor devices. \emph{Advanced materials (Deerfield Beach, Fla.)}
  \textbf{2014}, \emph{26}, 8163--8168\relax
\mciteBstWouldAddEndPuncttrue
\mciteSetBstMidEndSepPunct{\mcitedefaultmidpunct}
{\mcitedefaultendpunct}{\mcitedefaultseppunct}\relax
\EndOfBibitem
\bibitem[Madau{\ss} \latin{et~al.}(2018)Madau{\ss}, Zegkinoglou, {V{\'a}zquez
  Mui{\~n}os}, Choi, Kunze, Zhao, Naylor, Ernst, Pollmann, Ochedowski, Lebius,
  Benyagoub, Ban-d'Etat, Johnson, Djurabekova, {Roldan Cuenya}, and
  Schleberger]{Madau.2018}
Madau{\ss},~L.; Zegkinoglou,~I.; {V{\'a}zquez Mui{\~n}os},~H.; Choi,~Y.-W.;
  Kunze,~S.; Zhao,~M.-Q.; Naylor,~C.~H.; Ernst,~P.; Pollmann,~E.;
  Ochedowski,~O. \latin{et~al.}  Highly active single-layer MoS2 catalysts
  synthesized by swift heavy ion irradiation. \emph{Nanoscale} \textbf{2018},
  \emph{10}, 22908--22916\relax
\mciteBstWouldAddEndPuncttrue
\mciteSetBstMidEndSepPunct{\mcitedefaultmidpunct}
{\mcitedefaultendpunct}{\mcitedefaultseppunct}\relax
\EndOfBibitem
\bibitem[Feng \latin{et~al.}(2016)Feng, Graf, Liu, Ovchinnikov, Dumcenco,
  Heiranian, Nandigana, Aluru, Kis, and Radenovic]{Feng.2016}
Feng,~J.; Graf,~M.; Liu,~K.; Ovchinnikov,~D.; Dumcenco,~D.; Heiranian,~M.;
  Nandigana,~V.; Aluru,~N.~R.; Kis,~A.; Radenovic,~A. Single-layer MoS2
  nanopores as nanopower generators. \emph{Nature} \textbf{2016}, \emph{536},
  197--200\relax
\mciteBstWouldAddEndPuncttrue
\mciteSetBstMidEndSepPunct{\mcitedefaultmidpunct}
{\mcitedefaultendpunct}{\mcitedefaultseppunct}\relax
\EndOfBibitem
\bibitem[Heiranian \latin{et~al.}(2015)Heiranian, Farimani, and
  Aluru]{Heiranian.2015}
Heiranian,~M.; Farimani,~A.~B.; Aluru,~N.~R. Water desalination with a
  single-layer MoS2 nanopore. \emph{Nature communications} \textbf{2015},
  \emph{6}, 8616\relax
\mciteBstWouldAddEndPuncttrue
\mciteSetBstMidEndSepPunct{\mcitedefaultmidpunct}
{\mcitedefaultendpunct}{\mcitedefaultseppunct}\relax
\EndOfBibitem
\bibitem[Luan and Zhou(2018)Luan, and Zhou]{Luan.2018}
Luan,~B.; Zhou,~R. Single-File Protein Translocations through Graphene-MoS2
  Heterostructure Nanopores. \emph{The journal of physical chemistry letters}
  \textbf{2018}, \emph{9}, 3409--3415\relax
\mciteBstWouldAddEndPuncttrue
\mciteSetBstMidEndSepPunct{\mcitedefaultmidpunct}
{\mcitedefaultendpunct}{\mcitedefaultseppunct}\relax
\EndOfBibitem
\bibitem[Chen \latin{et~al.}(2018)Chen, Li, Zhang, Qiao, Tang, and
  Zhou]{Chen.2018}
Chen,~H.; Li,~L.; Zhang,~T.; Qiao,~Z.; Tang,~J.; Zhou,~J. Protein Translocation
  through a MoS 2 Nanopore:A Molecular Dynamics Study. \emph{The Journal of
  Physical Chemistry C} \textbf{2018}, \emph{122}, 2070--2080\relax
\mciteBstWouldAddEndPuncttrue
\mciteSetBstMidEndSepPunct{\mcitedefaultmidpunct}
{\mcitedefaultendpunct}{\mcitedefaultseppunct}\relax
\EndOfBibitem
\bibitem[Sarathy \latin{et~al.}(2018)Sarathy, Athreya, Varshney, and
  Leburton]{Sarathy.2018}
Sarathy,~A.; Athreya,~N.~B.; Varshney,~L.~R.; Leburton,~J.-P. Classification of
  Epigenetic Biomarkers with Atomically Thin Nanopores. \emph{The journal of
  physical chemistry letters} \textbf{2018}, \emph{9}, 5718--5725\relax
\mciteBstWouldAddEndPuncttrue
\mciteSetBstMidEndSepPunct{\mcitedefaultmidpunct}
{\mcitedefaultendpunct}{\mcitedefaultseppunct}\relax
\EndOfBibitem
\bibitem[Farimani \latin{et~al.}(2014)Farimani, Min, and Aluru]{Farimani.2014}
Farimani,~A.~B.; Min,~K.; Aluru,~N.~R. DNA base detection using a single-layer
  MoS2. \emph{ACS nano} \textbf{2014}, \emph{8}, 7914--7922\relax
\mciteBstWouldAddEndPuncttrue
\mciteSetBstMidEndSepPunct{\mcitedefaultmidpunct}
{\mcitedefaultendpunct}{\mcitedefaultseppunct}\relax
\EndOfBibitem
\bibitem[Feng \latin{et~al.}(2015)Feng, Liu, Bulushev, Khlybov, Dumcenco, Kis,
  and Radenovic]{Feng.2015}
Feng,~J.; Liu,~K.; Bulushev,~R.~D.; Khlybov,~S.; Dumcenco,~D.; Kis,~A.;
  Radenovic,~A. Identification of single nucleotides in MoS2 nanopores.
  \emph{Nature nanotechnology} \textbf{2015}, \emph{10}, 1070--1076\relax
\mciteBstWouldAddEndPuncttrue
\mciteSetBstMidEndSepPunct{\mcitedefaultmidpunct}
{\mcitedefaultendpunct}{\mcitedefaultseppunct}\relax
\EndOfBibitem
\bibitem[Smolyanitsky \latin{et~al.}(2016)Smolyanitsky, Yakobson, Wassenaar,
  Paulechka, and Kroenlein]{Smolyanitsky.2016}
Smolyanitsky,~A.; Yakobson,~B.~I.; Wassenaar,~T.~A.; Paulechka,~E.;
  Kroenlein,~K. A MoS2-Based Capacitive Displacement Sensor for DNA Sequencing.
  \emph{ACS nano} \textbf{2016}, \emph{10}, 9009--9016\relax
\mciteBstWouldAddEndPuncttrue
\mciteSetBstMidEndSepPunct{\mcitedefaultmidpunct}
{\mcitedefaultendpunct}{\mcitedefaultseppunct}\relax
\EndOfBibitem
\bibitem[Kou \latin{et~al.}(2016)Kou, Yao, Wu, Zhou, Lu, Wu, and Fan]{Kou.2016}
Kou,~J.; Yao,~J.; Wu,~L.; Zhou,~X.; Lu,~H.; Wu,~F.; Fan,~J. Nanoporous
  two-dimensional MoS2 membranes for fast saline solution purification.
  \emph{Physical chemistry chemical physics : PCCP} \textbf{2016}, \emph{18},
  22210--22216\relax
\mciteBstWouldAddEndPuncttrue
\mciteSetBstMidEndSepPunct{\mcitedefaultmidpunct}
{\mcitedefaultendpunct}{\mcitedefaultseppunct}\relax
\EndOfBibitem
\bibitem[Schleberger and Kotakoski(2018)Schleberger, and
  Kotakoski]{Schleberger.2018}
Schleberger,~M.; Kotakoski,~J. 2D Material Science: Defect Engineering by
  Particle Irradiation. \emph{Materials} \textbf{2018}, \emph{11},
  1885--1914\relax
\mciteBstWouldAddEndPuncttrue
\mciteSetBstMidEndSepPunct{\mcitedefaultmidpunct}
{\mcitedefaultendpunct}{\mcitedefaultseppunct}\relax
\EndOfBibitem
\bibitem[Komsa \latin{et~al.}(2012)Komsa, Kotakoski, Kurasch, Lehtinen, Kaiser,
  and Krasheninnikov]{Komsa.2012}
Komsa,~H.-P.; Kotakoski,~J.; Kurasch,~S.; Lehtinen,~O.; Kaiser,~U.;
  Krasheninnikov,~A.~V. Two-dimensional transition metal dichalcogenides under
  electron irradiation: defect production and doping. \emph{Physical review
  letters} \textbf{2012}, \emph{109}, 035503\relax
\mciteBstWouldAddEndPuncttrue
\mciteSetBstMidEndSepPunct{\mcitedefaultmidpunct}
{\mcitedefaultendpunct}{\mcitedefaultseppunct}\relax
\EndOfBibitem
\bibitem[Park \latin{et~al.}(2015)Park, Ryu, and Lee]{Park.2015}
Park,~H.~J.; Ryu,~G.~H.; Lee,~Z. Hole Defects on Two-Dimensional Materials
  Formed by Electron Beam Irradiation: Toward Nanopore Devices. \emph{Applied
  Microscopy} \textbf{2015}, \emph{45}, 107--114\relax
\mciteBstWouldAddEndPuncttrue
\mciteSetBstMidEndSepPunct{\mcitedefaultmidpunct}
{\mcitedefaultendpunct}{\mcitedefaultseppunct}\relax
\EndOfBibitem
\bibitem[Meyer \latin{et~al.}(2009)Meyer, Chuvilin, Algara-Siller, Biskupek,
  and Kaiser]{Meyer.2009}
Meyer,~J.~C.; Chuvilin,~A.; Algara-Siller,~G.; Biskupek,~J.; Kaiser,~U.
  Selective sputtering and atomic resolution imaging of atomically thin boron
  nitride membranes. \emph{Nano letters} \textbf{2009}, \emph{9},
  2683--2689\relax
\mciteBstWouldAddEndPuncttrue
\mciteSetBstMidEndSepPunct{\mcitedefaultmidpunct}
{\mcitedefaultendpunct}{\mcitedefaultseppunct}\relax
\EndOfBibitem
\bibitem[Feng \latin{et~al.}(2015)Feng, Liu, Graf, Lihter, Bulushev, Dumcenco,
  Alexander, Krasnozhon, Vuletic, Kis, and Radenovic]{Feng.2015b}
Feng,~J.; Liu,~K.; Graf,~M.; Lihter,~M.; Bulushev,~R.~D.; Dumcenco,~D.;
  Alexander,~D. T.~L.; Krasnozhon,~D.; Vuletic,~T.; Kis,~A. \latin{et~al.}
  Electrochemical Reaction in Single Layer MoS2: Nanopores Opened Atom by Atom.
  \emph{Nano letters} \textbf{2015}, \emph{15}, 3431--3438\relax
\mciteBstWouldAddEndPuncttrue
\mciteSetBstMidEndSepPunct{\mcitedefaultmidpunct}
{\mcitedefaultendpunct}{\mcitedefaultseppunct}\relax
\EndOfBibitem
\bibitem[Ghorbani-Asl \latin{et~al.}(2017)Ghorbani-Asl, Kretschmer, Spearot,
  and Krasheninnikov]{GhorbaniAsl.2017}
Ghorbani-Asl,~M.; Kretschmer,~S.; Spearot,~D.~E.; Krasheninnikov,~A.~V.
  Two-dimensional MoS 2 under ion irradiation: from controlled defect
  production to electronic structure engineering. \emph{2D Materials}
  \textbf{2017}, \emph{4}, 025078\relax
\mciteBstWouldAddEndPuncttrue
\mciteSetBstMidEndSepPunct{\mcitedefaultmidpunct}
{\mcitedefaultendpunct}{\mcitedefaultseppunct}\relax
\EndOfBibitem
\bibitem[Knust \latin{et~al.}(2017)Knust, Kreft, Hillmann, Meyer, Viefhues,
  Reimann, and Anselmetti]{Knust.2017}
Knust,~S.; Kreft,~D.; Hillmann,~R.; Meyer,~A.; Viefhues,~M.; Reimann,~P.;
  Anselmetti,~D. Measuring DNA Translocation Forces through MoS 2 -Nanopores
  with Optical Tweezers. \emph{Materials Today: Proceedings} \textbf{2017},
  \emph{4}, S168--S173\relax
\mciteBstWouldAddEndPuncttrue
\mciteSetBstMidEndSepPunct{\mcitedefaultmidpunct}
{\mcitedefaultendpunct}{\mcitedefaultseppunct}\relax
\EndOfBibitem
\bibitem[Madau{\ss} \latin{et~al.}(2017)Madau{\ss}, Ochedowski, Lebius,
  Ban-d'Etat, Naylor, Johnson, Kotakoski, and Schleberger]{Madau.2017}
Madau{\ss},~L.; Ochedowski,~O.; Lebius,~H.; Ban-d'Etat,~B.; Naylor,~C.~H.;
  Johnson,~A. T.~C.; Kotakoski,~J.; Schleberger,~M. Defect engineering of
  single- and few-layer MoS 2 by swift heavy ion irradiation. \emph{2D
  Materials} \textbf{2017}, \emph{4}, 015034\relax
\mciteBstWouldAddEndPuncttrue
\mciteSetBstMidEndSepPunct{\mcitedefaultmidpunct}
{\mcitedefaultendpunct}{\mcitedefaultseppunct}\relax
\EndOfBibitem
\bibitem[Gruber \latin{et~al.}(2016)Gruber, Salou, Bergen, {El Kharrazi},
  Lattouf, Grygiel, Wang, Benyagoub, Levavasseur, Rangama, Lebius, Ban-d'Etat,
  Schleberger, and Aumayr]{Gruber.2016}
Gruber,~E.; Salou,~P.; Bergen,~L.; {El Kharrazi},~M.; Lattouf,~E.; Grygiel,~C.;
  Wang,~Y.; Benyagoub,~A.; Levavasseur,~D.; Rangama,~J. \latin{et~al.}  Swift
  heavy ion irradiation of CaF2 - from grooves to hillocks in a single ion
  track. \emph{Journal of physics. Condensed matter : an Institute of Physics
  journal} \textbf{2016}, \emph{28}, 405001\relax
\mciteBstWouldAddEndPuncttrue
\mciteSetBstMidEndSepPunct{\mcitedefaultmidpunct}
{\mcitedefaultendpunct}{\mcitedefaultseppunct}\relax
\EndOfBibitem
\bibitem[Gruber \latin{et~al.}(2016)Gruber, Wilhelm, P{\'e}tuya, Smejkal,
  Kozubek, Hierzenberger, Bayer, Aldazabal, Kazansky, Libisch, Krasheninnikov,
  Schleberger, Facsko, Borisov, Arnau, and Aumayr]{Gruber.2016b}
Gruber,~E.; Wilhelm,~R.~A.; P{\'e}tuya,~R.; Smejkal,~V.; Kozubek,~R.;
  Hierzenberger,~A.; Bayer,~B.~C.; Aldazabal,~I.; Kazansky,~A.~K.; Libisch,~F.
  \latin{et~al.}  Ultrafast electronic response of graphene to a strong and
  localized electric field. \emph{Nature communications} \textbf{2016},
  \emph{7}, 13948\relax
\mciteBstWouldAddEndPuncttrue
\mciteSetBstMidEndSepPunct{\mcitedefaultmidpunct}
{\mcitedefaultendpunct}{\mcitedefaultseppunct}\relax
\EndOfBibitem
\bibitem[Ritter \latin{et~al.}(2012)Ritter, Wilhelm, Ginzel, Kowarik, Heller,
  El-Said, Papal{\'e}o, Rupp, {Crespo L{\'o}pez-Urrutia}, Ullrich, Facsko, and
  Aumayr]{Ritter.2012}
Ritter,~R.; Wilhelm,~R.~A.; Ginzel,~R.; Kowarik,~G.; Heller,~R.;
  El-Said,~A.~S.; Papal{\'e}o,~R.~M.; Rupp,~W.; {Crespo
  L{\'o}pez-Urrutia},~J.~R.; Ullrich,~J. \latin{et~al.}  Pit formation on
  poly(methyl methacrylate) due to ablation induced by individual slow highly
  charged ion impact. \emph{EPL (Europhysics Letters)} \textbf{2012},
  \emph{97}, 13001\relax
\mciteBstWouldAddEndPuncttrue
\mciteSetBstMidEndSepPunct{\mcitedefaultmidpunct}
{\mcitedefaultendpunct}{\mcitedefaultseppunct}\relax
\EndOfBibitem
\bibitem[Ritter \latin{et~al.}(2013)Ritter, Wilhelm, St{\"o}ger-Pollach,
  Heller, M{\"u}cklich, Werner, Vieker, Beyer, Facsko, G{\"o}lzh{\"a}user, and
  Aumayr]{Ritter.2013}
Ritter,~R.; Wilhelm,~R.~A.; St{\"o}ger-Pollach,~M.; Heller,~R.;
  M{\"u}cklich,~A.; Werner,~U.; Vieker,~H.; Beyer,~A.; Facsko,~S.;
  G{\"o}lzh{\"a}user,~A. \latin{et~al.}  Fabrication of nanopores in 1 nm thick
  carbon nanomembranes with slow highly charged ions. \emph{Applied Physics
  Letters} \textbf{2013}, \emph{102}, 063112\relax
\mciteBstWouldAddEndPuncttrue
\mciteSetBstMidEndSepPunct{\mcitedefaultmidpunct}
{\mcitedefaultendpunct}{\mcitedefaultseppunct}\relax
\EndOfBibitem
\bibitem[Hopster \latin{et~al.}(2014)Hopster, Kozubek, Ban-d'Etat, Guillous,
  Lebius, and Schleberger]{Hopster.2014}
Hopster,~J.; Kozubek,~R.; Ban-d'Etat,~B.; Guillous,~S.; Lebius,~H.;
  Schleberger,~M. Damage in graphene due to electronic excitation induced by
  highly charged ions. \emph{2D Materials} \textbf{2014}, \emph{1},
  011011\relax
\mciteBstWouldAddEndPuncttrue
\mciteSetBstMidEndSepPunct{\mcitedefaultmidpunct}
{\mcitedefaultendpunct}{\mcitedefaultseppunct}\relax
\EndOfBibitem
\bibitem[Kozubek \latin{et~al.}(2018)Kozubek, Ernst, Herbig, Michely, and
  Schleberger]{Kozubek.2018}
Kozubek,~R.; Ernst,~P.; Herbig,~C.; Michely,~T.; Schleberger,~M. Fabrication of
  Defective Single Layers of Hexagonal Boron Nitride on Various Supports for
  Potential Applications in Catalysis and DNA Sequencing. \emph{ACS Applied
  Nano Materials} \textbf{2018}, \emph{1}, 3765--3773\relax
\mciteBstWouldAddEndPuncttrue
\mciteSetBstMidEndSepPunct{\mcitedefaultmidpunct}
{\mcitedefaultendpunct}{\mcitedefaultseppunct}\relax
\EndOfBibitem
\bibitem[O'Brien \latin{et~al.}(2014)O'Brien, McEvoy, Hallam, Kim, Berner,
  Hanlon, Lee, Coleman, and Duesberg]{OBrien.2014}
O'Brien,~M.; McEvoy,~N.; Hallam,~T.; Kim,~H.-Y.; Berner,~N.~C.; Hanlon,~D.;
  Lee,~K.; Coleman,~J.~N.; Duesberg,~G.~S. Transition metal dichalcogenide
  growth via close proximity precursor supply. \emph{Scientific reports}
  \textbf{2014}, \emph{4}, 7374\relax
\mciteBstWouldAddEndPuncttrue
\mciteSetBstMidEndSepPunct{\mcitedefaultmidpunct}
{\mcitedefaultendpunct}{\mcitedefaultseppunct}\relax
\EndOfBibitem
\bibitem[Peters \latin{et~al.}(2009)Peters, Haake, Hopster, Sokolovsky, Wucher,
  and Schleberger]{Peters.2009}
Peters,~T.; Haake,~C.; Hopster,~J.; Sokolovsky,~V.; Wucher,~A.; Schleberger,~M.
  HICS: Highly charged ion collisions with surfaces. \emph{Nuclear Instruments
  and Methods in Physics Research Section B: Beam Interactions with Materials
  and Atoms} \textbf{2009}, \emph{267}, 687--690\relax
\mciteBstWouldAddEndPuncttrue
\mciteSetBstMidEndSepPunct{\mcitedefaultmidpunct}
{\mcitedefaultendpunct}{\mcitedefaultseppunct}\relax
\EndOfBibitem
\bibitem[Krivanek \latin{et~al.}(2010)Krivanek, Chisholm, Nicolosi, Pennycook,
  Corbin, Dellby, Murfitt, Own, Szilagyi, Oxley, Pantelides, and
  Pennycook]{Krivanek.2010}
Krivanek,~O.~L.; Chisholm,~M.~F.; Nicolosi,~V.; Pennycook,~T.~J.;
  Corbin,~G.~J.; Dellby,~N.; Murfitt,~M.~F.; Own,~C.~S.; Szilagyi,~Z.~S.;
  Oxley,~M.~P. \latin{et~al.}  Atom-by-atom structural and chemical analysis by
  annular dark-field electron microscopy. \emph{Nature} \textbf{2010},
  \emph{464}, 571--574\relax
\mciteBstWouldAddEndPuncttrue
\mciteSetBstMidEndSepPunct{\mcitedefaultmidpunct}
{\mcitedefaultendpunct}{\mcitedefaultseppunct}\relax
\EndOfBibitem
\bibitem[Hopster \latin{et~al.}(2013)Hopster, Kozubek, Kr{\"a}mer, Sokolovsky,
  and Schleberger]{Hopster.2013}
Hopster,~J.; Kozubek,~R.; Kr{\"a}mer,~J.; Sokolovsky,~V.; Schleberger,~M.
  Ultra-thin MoS2 irradiated with highly charged ions. \emph{Nuclear
  Instruments and Methods in Physics Research Section B: Beam Interactions with
  Materials and Atoms} \textbf{2013}, \emph{317}, 165--169\relax
\mciteBstWouldAddEndPuncttrue
\mciteSetBstMidEndSepPunct{\mcitedefaultmidpunct}
{\mcitedefaultendpunct}{\mcitedefaultseppunct}\relax
\EndOfBibitem
\bibitem[Li \latin{et~al.}(2015)Li, Wang, Zhang, Zhao, Duan, and Xue]{Li.2015}
Li,~W.; Wang,~X.; Zhang,~X.; Zhao,~S.; Duan,~H.; Xue,~J. Mechanism of the
  defect formation in supported graphene by energetic heavy ion irradiation:
  the substrate effect. \emph{Scientific reports} \textbf{2015}, \emph{5},
  9935\relax
\mciteBstWouldAddEndPuncttrue
\mciteSetBstMidEndSepPunct{\mcitedefaultmidpunct}
{\mcitedefaultendpunct}{\mcitedefaultseppunct}\relax
\EndOfBibitem
\bibitem[Zhao and Xue(2015)Zhao, and Xue]{Zhao.2015}
Zhao,~S.; Xue,~J. Modification of graphene supported on SiO 2 substrate with
  swift heavy ions from atomistic simulation point. \emph{Carbon}
  \textbf{2015}, \emph{93}, 169--179\relax
\mciteBstWouldAddEndPuncttrue
\mciteSetBstMidEndSepPunct{\mcitedefaultmidpunct}
{\mcitedefaultendpunct}{\mcitedefaultseppunct}\relax
\EndOfBibitem
\bibitem[Kretschmer \latin{et~al.}(2018)Kretschmer, Maslov, Ghaderzadeh,
  Ghorbani-Asl, Hlawacek, and Krasheninnikov]{Kretschmer.2018}
Kretschmer,~S.; Maslov,~M.; Ghaderzadeh,~S.; Ghorbani-Asl,~M.; Hlawacek,~G.;
  Krasheninnikov,~A.~V. Supported Two-Dimensional Materials under Ion
  Irradiation: The Substrate Governs Defect Production. \emph{ACS applied
  materials {\&} interfaces} \textbf{2018}, \emph{10}, 30827--30836\relax
\mciteBstWouldAddEndPuncttrue
\mciteSetBstMidEndSepPunct{\mcitedefaultmidpunct}
{\mcitedefaultendpunct}{\mcitedefaultseppunct}\relax
\EndOfBibitem
\bibitem[Schiwietz \latin{et~al.}(1995)Schiwietz, Briere, Schneider, McDonald,
  and Cunningham]{Schiwietz.1995}
Schiwietz,~G.; Briere,~M.; Schneider,~D.; McDonald,~J.; Cunningham,~C.
  Measurement of negative-ion and -cluster sputtering with highly-charged heavy
  ions. \emph{Nuclear Instruments and Methods in Physics Research Section B:
  Beam Interactions with Materials and Atoms} \textbf{1995}, \emph{100},
  47--54\relax
\mciteBstWouldAddEndPuncttrue
\mciteSetBstMidEndSepPunct{\mcitedefaultmidpunct}
{\mcitedefaultendpunct}{\mcitedefaultseppunct}\relax
\EndOfBibitem
\bibitem[Wilhelm and M{\"o}ller(2016)Wilhelm, and M{\"o}ller]{Wilhelm.2016}
Wilhelm,~R.~A.; M{\"o}ller,~W. Charge-state-dependent energy loss of slow ions.
  II. Statistical atom model. \emph{Physical Review A} \textbf{2016},
  \emph{93}, 1\relax
\mciteBstWouldAddEndPuncttrue
\mciteSetBstMidEndSepPunct{\mcitedefaultmidpunct}
{\mcitedefaultendpunct}{\mcitedefaultseppunct}\relax
\EndOfBibitem
\bibitem[Plimpton(1995)]{Plimpton.1995}
Plimpton,~S. Fast Parallel Algorithms for Short-Range Molecular Dynamics.
  \emph{Journal of Computational Physics} \textbf{1995}, \emph{117},
  1--19\relax
\mciteBstWouldAddEndPuncttrue
\mciteSetBstMidEndSepPunct{\mcitedefaultmidpunct}
{\mcitedefaultendpunct}{\mcitedefaultseppunct}\relax
\EndOfBibitem
\bibitem[Ziegler \latin{et~al.}(2010)Ziegler, Ziegler, and
  Biersack]{Ziegler.2010}
Ziegler,~J.~F.; Ziegler,~M.~D.; Biersack,~J.~P. SRIM -- The stopping and range
  of ions in matter (2010). \emph{Nuclear Instruments and Methods in Physics
  Research Section B: Beam Interactions with Materials and Atoms}
  \textbf{2010}, \emph{268}, 1818--1823\relax
\mciteBstWouldAddEndPuncttrue
\mciteSetBstMidEndSepPunct{\mcitedefaultmidpunct}
{\mcitedefaultendpunct}{\mcitedefaultseppunct}\relax
\EndOfBibitem
\bibitem[Wilhelm \latin{et~al.}(2017)Wilhelm, Gruber, Schwestka, Kozubek,
  Madeira, Marques, Kobus, Krasheninnikov, Schleberger, and
  Aumayr]{Wilhelm.2017}
Wilhelm,~R.~A.; Gruber,~E.; Schwestka,~J.; Kozubek,~R.; Madeira,~T.~I.;
  Marques,~J.~P.; Kobus,~J.; Krasheninnikov,~A.~V.; Schleberger,~M.; Aumayr,~F.
  Interatomic Coulombic Decay: The Mechanism for Rapid Deexcitation of Hollow
  Atoms. \emph{Physical review letters} \textbf{2017}, \emph{119}, 103401\relax
\mciteBstWouldAddEndPuncttrue
\mciteSetBstMidEndSepPunct{\mcitedefaultmidpunct}
{\mcitedefaultendpunct}{\mcitedefaultseppunct}\relax
\EndOfBibitem
\bibitem[Lin \latin{et~al.}(2016)Lin, Carvalho, Kahn, Lv, Rao, Terrones,
  Pimenta, and Terrones]{Lin.2016}
Lin,~Z.; Carvalho,~B.~R.; Kahn,~E.; Lv,~R.; Rao,~R.; Terrones,~H.;
  Pimenta,~M.~A.; Terrones,~M. Defect engineering of two-dimensional transition
  metal dichalcogenides. \emph{2D Materials} \textbf{2016}, \emph{3},
  022002\relax
\mciteBstWouldAddEndPuncttrue
\mciteSetBstMidEndSepPunct{\mcitedefaultmidpunct}
{\mcitedefaultendpunct}{\mcitedefaultseppunct}\relax
\EndOfBibitem
\bibitem[Komsa \latin{et~al.}(2013)Komsa, Kurasch, Lehtinen, Kaiser, and
  Krasheninnikov]{Komsa.2013}
Komsa,~H.-P.; Kurasch,~S.; Lehtinen,~O.; Kaiser,~U.; Krasheninnikov,~A.~V. From
  point to extended defects in two-dimensional MoS 2 : Evolution of atomic
  structure under electron irradiation. \emph{Physical Review B} \textbf{2013},
  \emph{88}\relax
\mciteBstWouldAddEndPuncttrue
\mciteSetBstMidEndSepPunct{\mcitedefaultmidpunct}
{\mcitedefaultendpunct}{\mcitedefaultseppunct}\relax
\EndOfBibitem
\bibitem[Wang \latin{et~al.}(2017)Wang, Li, Sawada, Allen, Kirkland, Grossman,
  and Warner]{Wang.2017}
Wang,~S.; Li,~H.; Sawada,~H.; Allen,~C.~S.; Kirkland,~A.~I.; Grossman,~J.~C.;
  Warner,~J.~H. Atomic structure and formation mechanism of sub-nanometer pores
  in 2D monolayer MoS2. \emph{Nanoscale} \textbf{2017}, \emph{9},
  6417--6426\relax
\mciteBstWouldAddEndPuncttrue
\mciteSetBstMidEndSepPunct{\mcitedefaultmidpunct}
{\mcitedefaultendpunct}{\mcitedefaultseppunct}\relax
\EndOfBibitem
\bibitem[Zan \latin{et~al.}(2013)Zan, Ramasse, Jalil, Georgiou, Bangert, and
  Novoselov]{Zan.2013}
Zan,~R.; Ramasse,~Q.~M.; Jalil,~R.; Georgiou,~T.; Bangert,~U.; Novoselov,~K.~S.
  Control of radiation damage in MoS(2) by graphene encapsulation. \emph{ACS
  nano} \textbf{2013}, \emph{7}, 10167--10174\relax
\mciteBstWouldAddEndPuncttrue
\mciteSetBstMidEndSepPunct{\mcitedefaultmidpunct}
{\mcitedefaultendpunct}{\mcitedefaultseppunct}\relax
\EndOfBibitem
\bibitem[Aumayr \latin{et~al.}(2011)Aumayr, Facsko, El-Said, Trautmann, and
  Schleberger]{Aumayr.2011}
Aumayr,~F.; Facsko,~S.; El-Said,~A.~S.; Trautmann,~C.; Schleberger,~M. Single
  ion induced surface nanostructures: a comparison between slow highly charged
  and swift heavy ions. \emph{Journal of physics. Condensed matter : an
  Institute of Physics journal} \textbf{2011}, \emph{23}, 393001\relax
\mciteBstWouldAddEndPuncttrue
\mciteSetBstMidEndSepPunct{\mcitedefaultmidpunct}
{\mcitedefaultendpunct}{\mcitedefaultseppunct}\relax
\EndOfBibitem
\bibitem[Wilhelm \latin{et~al.}(2015)Wilhelm, Gruber, Ritter, Heller, Beyer,
  Turchanin, Klingner, H{\"u}bner, St{\"o}ger-Pollach, Vieker, Hlawacek,
  G{\"o}lzh{\"a}user, Facsko, and Aumayr]{Wilhelm.2015}
Wilhelm,~R.~A.; Gruber,~E.; Ritter,~R.; Heller,~R.; Beyer,~A.; Turchanin,~A.;
  Klingner,~N.; H{\"u}bner,~R.; St{\"o}ger-Pollach,~M.; Vieker,~H.
  \latin{et~al.}  Threshold and efficiency for perforation of 1 nm thick carbon
  nanomembranes with slow highly charged ions. \emph{2D Materials}
  \textbf{2015}, \emph{2}, 035009\relax
\mciteBstWouldAddEndPuncttrue
\mciteSetBstMidEndSepPunct{\mcitedefaultmidpunct}
{\mcitedefaultendpunct}{\mcitedefaultseppunct}\relax
\EndOfBibitem
\end{mcitethebibliography}
\end{document}